\documentclass[10pt,conference]{IEEEtran}
\usepackage{mathptmx} % This is Times font
\usepackage{amsmath,amssymb,amsfonts}
\usepackage{comment}
\usepackage{fancyhdr}
\usepackage[normalem]{ulem}
\usepackage[hyphens]{url}
\usepackage{cite}
\usepackage[final]{microtype}
\usepackage[keeplastbox]{flushend}
\usepackage[bookmarks=true,breaklinks=true,letterpaper=true,colorlinks,linkcolor=black,citecolor=blue,urlcolor=black]{hyperref}

\usepackage{ifthen}
\usepackage{xcolor}
\usepackage[export]{adjustbox}% <-- added
\usepackage{siunitx}
\usepackage{enumitem}
\usepackage{multicol}
\usepackage{tikz}
\usepackage[utf8]{inputenc}
\usepackage{xcolor,soul}
\usepackage{array,multirow,graphicx}
\usepackage{float}
\usepackage{subfig}
\usepackage{footnote}
\usepackage{pifont}
\usepackage[linesnumbered,algoruled,boxed,lined]{algorithm2e}
\usepackage{array}

\def\BibTeX{{\rm B\kern-.05em{\sc i\kern-.025em b}\kern-.08em
    T\kern-.1667em\lower.7ex\hbox{E}\kern-.125emX}}
    
\title{\vspace{-4mm}Quantum Circuit Resizing\vspace{-5mm}} 
\newcommand{\squishlist}{
	\begin{list}{$\bullet$}
		{ \setlength{\itemsep}{0pt}      \setlength{\parsep}{0pt}
			\setlength{\topsep}{0.5pt}       \setlength{\partopsep}{0pt}
			\setlength{\listparindent}{-2pt}
			\setlength{\itemindent}{-5pt}
			\setlength{\leftmargin}{1em} \setlength{\labelwidth}{0em}
			\setlength{\labelsep}{0.5em} } }
	
	\newcommand{\squishend}{
\end{list}  }

\newcommand{\dashlist}{
	\begin{list}{--}
		{ \setlength{\itemsep}{0pt}      \setlength{\parsep}{0pt}
			\setlength{\topsep}{0.5pt}       \setlength{\partopsep}{0pt}
			\setlength{\listparindent}{-2pt}
			\setlength{\itemindent}{-5pt}
			\setlength{\leftmargin}{1em} \setlength{\labelwidth}{0em}
			\setlength{\labelsep}{0.5em} } }
	
	\newcommand{\listend}{
\end{list}  }

\title{Quantum Circuit Resizing} 
\author{
\IEEEauthorblockN{Movahhed Sadeghi\IEEEauthorrefmark{1}}
\IEEEauthorblockA{The Pennsylvania State University,\\Department of\\Computer Science and Engineering\\
Email: mus883@psu.edu}\\   
\IEEEauthorrefmark{1}corresponding Author
\and
\IEEEauthorblockN{Soheil Khadirsharbiyani}
\IEEEauthorblockA{The Pennsylvania State University,\\Department of\\Computer Science and Engineering\\
Email: szk921@psu.edu}\\   
\and
\IEEEauthorblockN{Mahmut Taylan Kandemir}
\IEEEauthorblockA{The Pennsylvania State University,\\Department of\\Computer Science and Engineering\\
Email: mtk2@psu.edu}\\   

}

\begin{document}

\maketitle
\thispagestyle{plain}
\pagestyle{plain}

%%%%%% -- PAPER CONTENT STARTS-- %%%%%%%%

%%%%%

\begin{abstract}
Existing quantum systems provide very limited physical qubit counts, trying to execute a quantum algorithm/circuit on them that have a higher number of logical qubits than physically available lead to a compile-time error. Given that it is unrealistic to expect existing quantum systems to provide, in near future, sufficient number of qubits that can accommodate large circuit, there is a pressing need to explore strategies that can somehow execute large circuits on small systems.  

In this paper, first, we perform an analysis to identify the  qubits that are most suitable for circuit resizing. Our results reveal that, in most quantum programs, there exist qubits that can be reused mid-program to serially/sequentially execute the circuit employing fewer qubits. Motivated by this observation, we  design, implement and evaluate a compiler-based approach that i) identifies the qubits that can be most beneficial for serial circuit execution; ii) selects those qubits to reuse at each step of execution for size minimization of the circuit; and iii) minimizes Middle Measurement (MM) delays due to impractical implementation of shots to improve the circuit reliability. Furthermore, since our approach intends to execute the circuits sequentially, the crosstalk errors can also be optimized as a result of the reduced number of concurrent gates.

The experimental  results  indicate that our proposed approach can (i) execute large circuits that initially cannot fit into small circuits, on small quantum hardware, and (ii) can significantly improve the PST of the results by 2.1X when both original and our serialized programs can fit into the target quantum hardware.   

%Motivated by this observation, in this work, we present and evaluate a strategy for reducing the size of quantum circuits on NISQ systems that feature middle measurement (MM) and middle reset (MR) gates. Normally, to run a quantum program on any NISQ machine, the number of logical qubits should not exceed the number of physical qubits available in the quantum hardware. This serves as an extra incentive for the present hardware developer to enhance the system's size. As a result, the small accessible systems are now just useful for testing and demonstrating certain small quantum circuit instances. We discover that some larger circuits may fit comfortably on these platforms while achieving reliable output using our minimization approach. Additionally, by serializing the operation, we can minimize the number of SWAPs, %leading to significant PST improvement. Our approach benefits from MM and MR gates to expand the possibilities for compiling quantum circuits. Our results indicate that we can achieve up to YYY qubit requirement reduction while improving the reliability by a factor of ZZZ.

\end{abstract} 

%%%%%

\section{Introduction} \label{sec:intro} 
Quantum computers are introduced to enhance the computational capacity for complex problems such as machine  learning~\cite{biamonte2017quantum} and chemistry  simulation~\cite{kandala2017hardware}. Over the last decade, several vendors unveiled their quantum hardware in an effort to benefit from quantum phenomena to execute a quantum program on real systems. Currently, vendors like Google, IBM, and Intel support up to 72, 127, and 49 qubits (quantum bits)~\cite{google,Intel49,IBM}, respectively. However, due to errors introduced in the real implementation of qubits, such as coherence and gate errors, existing quantum computers are highly unreliable. To solve this issue, Quantum Error Correction Codes (QEC)  algorithms~\cite{fowler2012surface,hu2019quantum,campagne2020quantum,ma2020error} have been introduced, but they typically require 10-100 additional qubits to create a single fault-tolerant qubit.   
                             
Due to this enormous extra qubit requirements brought by QEC, which is impractical considering the size of the current systems, Noisy Intermediate Scale Quantum Computers (NISQ)  \cite{fowler2012surface} have been developed to execute small-to-medium circuits on current hardware while aiming to minimize the error rate by different techniques. Several works targeting NISQ machines have been introduced to minimize different errors, aiming to improve the reliability of the system. While some works such as~\cite{liu2021qucloud} try to execute a given quantum program on the most reliable set of qubits available in the target hardware, others~\cite{liu2021qucloud,tannu2019not,li2019tackling,zhang2020depth} aim to minimize gate errors by optimizing the number of SWAP operations (SWAP operations are used to transfer the content of one qubit to another when the qubits involved in a 2-qubit operation do not have a direct physical connection between them). Despite these efforts, the problem of executing a large quantum circuit with lots of logical qubits on existing small quantum systems with only a few physical qubits seems to be one of the biggest challenges in quantum computing. 

Till recently, there had been no mechanism to solve this problem successfully. However, quantum vendors have recently introduced the {\em Middle Reset} (MR) and {\em Middle Measurement} (MM) gates, which can be utilized to {\em resize} quantum circuits during the execution. By utilizing MR/MM, theoretically, famous quantum algorithms like  Bernstein-Vazirani~\cite{bernstein1997quantum} with any qubit count can be executed only using 2 qubits \cite{middleresetIBM}. Additionally, when running on 2 qubits, no SWAP operations is needed since the 2 physical qubits to which the program is assigned usually have a connection between them, eliminating the gate error due to SWAP operations in the process. Although this approach can be highly effective in improving the system's reliability when the qubits we want to reset are correctly identified, to our knowledge, there are no prior works in this area that exploit the potential of these gates in an automated fashion. 

In this paper, first, we perform an analysis to identify the  qubits that are most suitable for circuit resizing. Our results reveal that, in most quantum programs, there exist qubits that can be reused mid-program to {\em serially execute} the circuit employing fewer qubits. Motivated by this observation, we  design, implement and evaluate a compiler-based approach that i) identifies the qubits that can be most beneficial for serial circuit execution; ii) selects those qubits to reuse at each step of execution for size minimization of the circuit\footnote{When it leads to no confusion, we will use the terms "serializability", "sequential/serial execution", "circuit reduction", and "circuit resizing", interchangeably.}; and iii) minimizes Middle Measurement (MM) delays due to impractical implementation of shots\footnote{Numerous executions, known as "shots", are often carried out in order to obtain the chance of getting the right answer from quantum hardware.} to improve the circuit reliability. 
%; iv) in addition to minimizing the qubit requirement for a circuit through a novel algorithm that can finish its task in polynomial time. 
Furthermore, since our approach intends to execute the circuits sequentially, the crosstalk errors can also be optimized as a result of the reduced number of concurrent gates. To summarize, in this paper, we make the following {\bf main contributions}: 

\squishlist
    \item We observe that there is only one constraint that need to be satisfied in a quantum circuit for the circuit to be "serially executable". More detail about this constraint is given in Section~\ref{sec:design}.
    \item Based on this observation, we present an algorithm that selects qubits in a way that the serial execution opportunity is maximized; hence, the size of the resulting circuit is minimal.   
    \item We present a proof demonstrating that our proposed approach really minimizes the number of qubits needed to execute a quantum program (i.e., it completely serializes it). Consequently,  we avoid system size compilation error whenever it is possible to do so. 
    \item We observe that the current concept of {\em shot} does not fully work with our proposal and leads to coherence errors in the results. To solve it, we present a new concept/feature in quantum systems called {\em  iteration}, which leads to highly reliable results.  
    \item We present experimental evidence showing the  effectiveness of our proposed approach. The experimental  results  indicate that our proposed approach can (i) execute large circuits that initially cannot fit into small circuits, on small quantum hardware, and (ii) can significantly improve the PST of the results by 2.1x when both original and our serialized programs can fit into the target quantum hardware.    
\squishend  

The remainder of this paper is organized as follows. In Section~\ref{sec:background}, we give a background on  quantum computing covering quantum computing basics,  currently available quantum hardware, and MM/MR gates. In Section~\ref{sec:related}, we go over the prior works relevant to this study and explain their shortcomings. In  Section~\ref{sec:motiv}, we motivate our work by characterizing the problem, and in  Section~\ref{sec:design}, we  explain the technical details of our proposed approach to circuit minimization/serialization. In Section~\ref{sec:methodology}, we present our evaluation setup and describe the workloads as well as our evaluation methodology. The performance and sensitivity results are presented, respectively, Sections~\ref{sec:result} and~\ref{sec:sens}. In Section~\ref{sec:conclusion}, we give a summary of our major conclusions, and finally, discuss future research directions in Section~\ref{sec:future}. 

%%%%%

%%%%%

\section{Background and Related Work} \label{sec:background}
In this section, we give an overview of quantum computing and  discuss representative quantum systems that are currently available and their salient features. 

\subsection{Quantum Computing Basics}
\label{back-qucomputation}
Quantum computation is based on {\em qubits}, as opposed to classical computing, which is based on bits. Compared to a classical bit which represents a value of $0$ or $1$, a qubit is represented as a {\em vector} that holds a "state" {\em between} $0$ and $1$, which is defined  as follows: $|\varphi\rangle = \alpha |0\rangle + \beta |1\rangle$. This leads to an exponential growth in state space in terms of the number of qubits~\cite{shor1999polynomial}. For example, having two qubits gives us a state space of $|\varphi\rangle = \alpha_{00} |00\rangle + \alpha_{01} |01\rangle +\alpha_{10} |10\rangle +\alpha_{11} |11\rangle.$ Consequently, algorithms whose state/search space grows exponentially can potentially benefit from quantum computing by operating on qubits and linearizing their state space~\cite{shor1999polynomial}. 

A {\em quantum gate} is the basic building block of a {\em quantum program}. It typically operates on a small number of qubits. Example quantum gates include SWAP gate, NOT-gate square root, Controlled-NOT gate (C-NOT) and other controlled gates. It is to be emphasized that a quantum gate that operates on multiple qubits can execute only in the presence of a {\em direct link} between the involved qubits.\footnote{We distinguish between "logical qubits" and "physical qubits"; the former is an abstract qubit in a quantum program or quantum circuit, whereas the latter represents a physical device (which can be implemented in various ways/technologies) that acts as a two-state quantum system.} In reality, qubits are prone to a variety of errors, such as coherence error, gate error, and crosstalk. Qubits can only keep their state for a finite amount of time, which leads to coherence error.
This error increases exponentially over time by a factor of $\frac{t}{T1}$ or $\frac{t}{T2}$, where $T1$ is the time it takes to transition from a state of $|1\rangle$ to $|0\rangle$ and $T2$ is the time it takes to transition from a middle state to a $|0\rangle$ state for a qubit, according to~\cite{murali2020software}. Gate error happens because of the operations executing on qubits, and crosstalk happens due to the interaction between different qubits during concurrently running operations. Additionally, during the measurement, there is another type of error, called readout error, that affects the reliability of the outputs. Note that T1, T2, gate error, and readout error are the system's characteristics and are different across different systems.

\begin{figure}[t!]
    \centering
    \vspace{-5mm}
    \includegraphics[width=0.7\columnwidth]{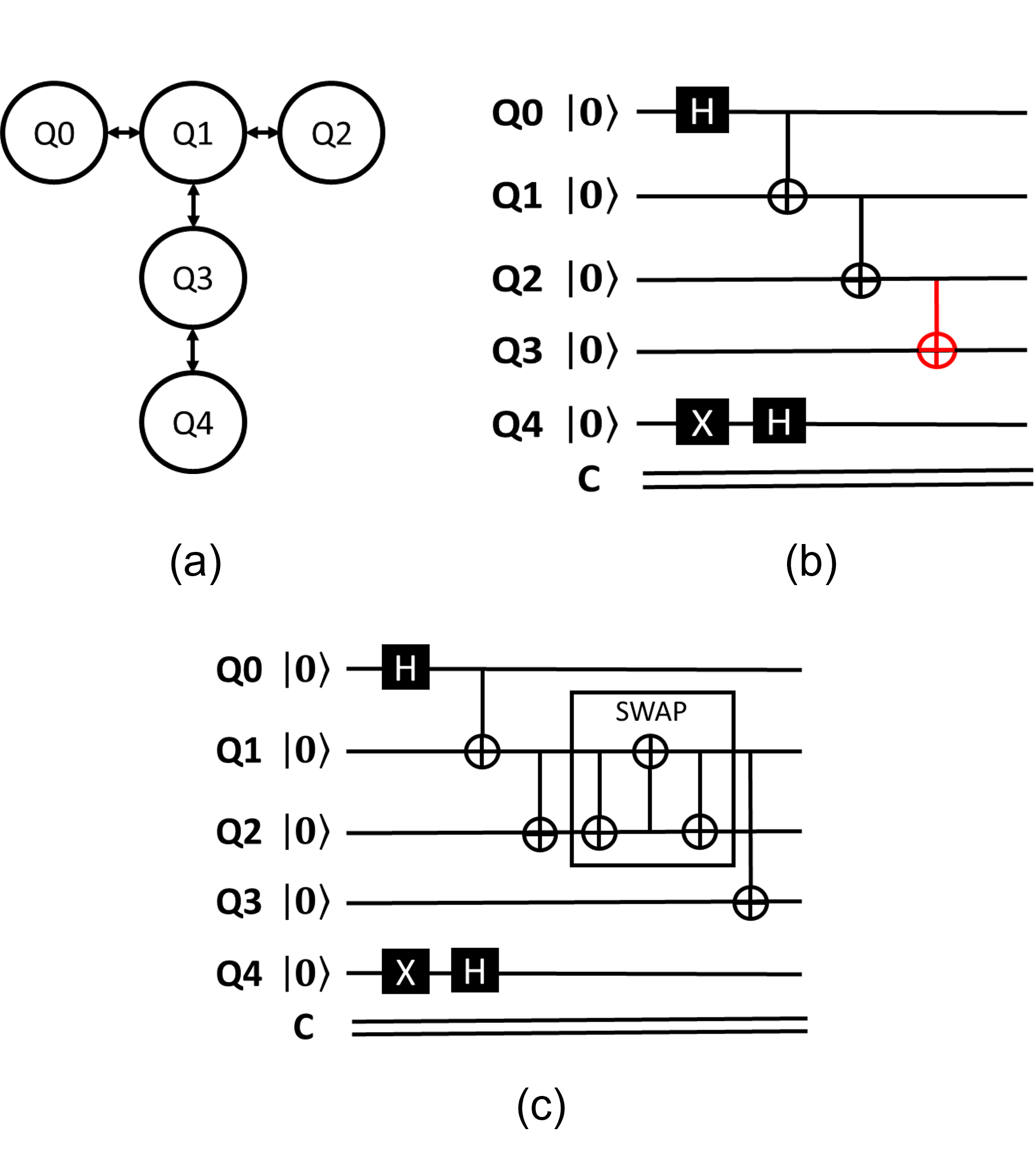}\label{fig:backgroundSWAP}%\par
    \vspace{-5pt}
    \caption{(a) Architecture of the system, (b) Sample circuit with no SWAP operation, and (c) The same circuit with the addition of a SWAP operation.}
    \label{fig:swap}
    \vspace{-6mm}
\end{figure}

To solve the reliability concerns, quantum systems require quantum error correction (QEC) to ensure the correctness of the results. However, currently, QEC codes need the addition of numerous "extra" qubits to ensure the reliability of a single qubit, which is impractical given the limited scale of present systems. This brings about the  NISQ~\cite{preskill2018quantum} era, which permits the execution of small-to-medium size circuits on quantum systems without any QEC techniques. In NISQ (Noisy Intermediate-Scale Quantum) devices, the connections between qubits are limited, for reliability purposes, to avoid the requirement of QEC~\cite{tannu2019not}. Instead, NISQ machines rely heavily on SWAP operations~\cite{Cambridge2007Book} -- gate operations that swap the state of two linked/neighboring  qubits\footnote{Neighboring qubits are qubits that are connected to each other via a direct physical link.}, to perform an operation between two qubits that are not  adjacent. More specifically, when an operation is anticipated between two qubits that are not neighbors, one of the qubits would swap across links until it becomes a neighbor of the other qubit~\cite{murali2020software}. Following that, the original gate operation between them can finally take place. 
Fig.~\ref{fig:swap} shows an example of how SWAP operations are added to make a circuit executable. Fig.~\ref{fig:swap}-a shows the architecture of the system. Based on this architecture, for the circuit shown in Fig.~\ref{fig:swap}-b, there is no direct connection between Q2 and Q3 for the red CNOT to be executed. Therefore, we switch the content of Q1 and Q2 by using the SWAP operation so that we can run the CNOT operation mentioned above. After adding the SWAP, the CNOT operation is performed between Q1 and Q3, generating the final result. To sustain reliability, as the scale of quantum systems grows, the number of links across qubits reduces. Current IBMQ systems, for example, include one to three links per qubit, with the bulk of qubits connected to only two links~\cite{ibmq}. This in turn causes an increase in the number of swap gates required for a circuit to execute on them.

Current NISQ systems operate in a QAOA (Quantum Approximate Optimization Algorithm)~\cite{farhi2014quantum} fashion. QAOA is a paradigm that combines classical computers and NISQ systems~\cite{farhi2014quantum}. More specifically, a quantum program compiles into either a quantum circuit or a batch of quantum circuits. At the end of the execution of each circuit, the resulting qubits (output) are measured and their measured values are stored in the "classical computer memory". Subsequently, the qubits are reset to a state of $|0\rangle$ for the next circuit to execute. Note that due to the "probabilistic nature" of quantum computing, the results of various executions may differ. As a result, several runs, also known as "shots," are often performed to determine the probability of having the correct answer.

%%%%%
\subsection{Different Types of Dependency in QC} 
%%%%%%%%%%%%%%%%
Qiskit provides a circuit directed acyclic graph DAG for users, in which roots and leaves represent qubits and other nodes represent gate operations. Additionally, the edges represent the qubits used by each gates operations as shown in the example of Fig.~\ref{fig:BVN}.
The use cases of DAG include (but are not limited to) providing order of gate operations, dependencies between qubits, parallel operations at each stage, and critical depth of the circuit.  
Programmer uses for DAG include computation of circuit execution speed, locating parallel CNOT operations to avoid cross-talk, locating the circuit's critical CNOT path for different optimizations, etc.
While DAG provides the dependencies between operations, it does not differentiate between false and true dependencies. False dependencies exist in DAG due to the sequence of operations but do not influence the subsequent qubit/operation, meaning that changing the sequence does not affect the outcome. However, true dependencies can affect the final result of the system if the dependency is not satisfied (order of operations change). In Fig.~\ref{fig:BVN}, for example, the CNOT operation between $q_0$ and $q_5$ has no influence on the value of $q_1$ following its CNOT operation with $q_5$, but in the DAG, they are dependent due to the sequence of the program. 
\begin{figure}[t!]
    \centering
    \includegraphics[width=.9\columnwidth]{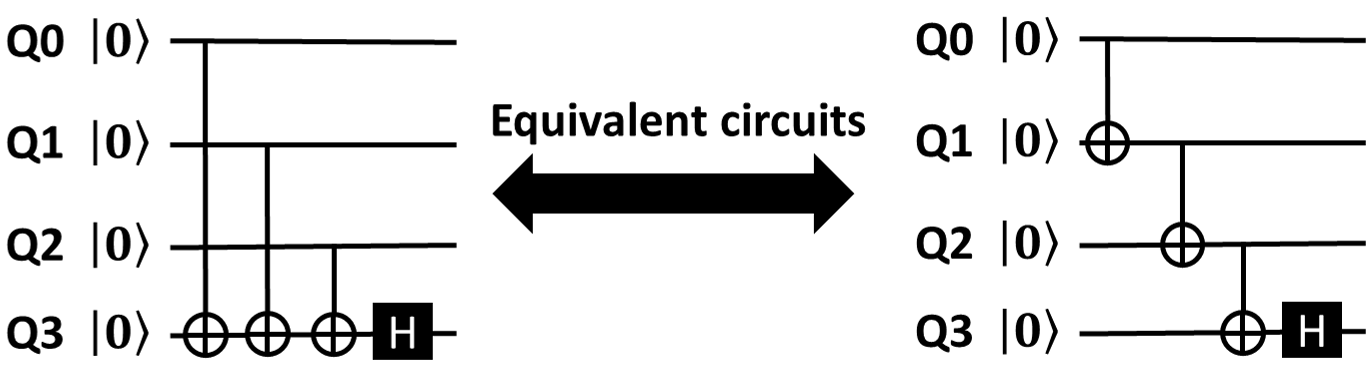}\label{fig:False dependency}%\par
    \vspace{-5pt}
    \caption{Two equivalent circuits obtain from ignoring false dependencies, However they both can be minimized to 2 qubits}
    \label{fig:FD}
    \vspace{-6mm}
\end{figure}
Fig.~\ref{fig:FD} depicts an example in which two circuits are equivalent and can transform to each other by taking advantage of false dependencies. False dependencies can provide the opportunity to change the order of the operations and transform circuits~\cite{li2022paulihedral}. Attaining these false dependencies is easy during the early stages of compilation but becomes impractical to detect during the execution of the operations.

%%%%%%%%%%%%%%%
\subsection{New Features: MM and MR} 
Starting in 2020, IBM Quantum systems (IBMQ) started to  gradually include {\em Middle Measurement} (MM) and {\em Middle Reset} (MR) gates into their quantum systems~\cite{middleresetIBM}, aiming to provide {\em qubit reuse} during the course of program execution. To our knowledge, all IBMQ quantum systems currently support these features. In the early days of these gates' debut, IBMQ provided an instance of the  Bernstein-Vazirani~\cite{bernstein1997quantum,du2001implementation} circuit (a 5-qubit version is shown in Fig.~\ref{fig:BVN}) to demonstrate the possibility of serial execution of quantum circuits for improved reliability~\cite{middleresetIBM}. The presented results demonstrate that, by employing the MM and MR gates, some quantum circuits (but not all) can be executed "serially/sequentially" using a smaller number of qubits. An example is illustrated in Fig.~\ref{fig:Serial}. Still, to our knowledge, there is {\em no} automated approach to  downsize/serialize a given quantum circuit by taking advantage of MM/MR. As a result, currently, if the circuit's number of qubits exceeds the target system's  physical qubit count, the compiler generates an "error" and does {\em not} execute the circuit on the target system, whereas, with a proper/careful use of MM and MR, depending on the circuit at hand, the compiler can still be able to execute the circuit successfully. This paper proposes a compiler-based "circuit resizing/serialization" approach that {\em automates} the disciplined use of MM/MR so that a given quantum circuit can execute on fewer qubits -- in a serial fashion -- without any compile-time error.   

%Still, to our knowledge, no attempt has been made to downsize suitable quantum circuits for execution on quantum systems, and no existing framework is available to benefit from this feature. This means that, if the circuit's number of qubits exceeds the system's number of qubits, the compiler generates an error and does not execute the circuit on the target system  (Fig.~\ref{fig:Error}), whereas, with MM and MR, the compiler can still be able to run the circuit. 

%%%%%

% \begin{Fig.}[ht!]
% \center
% \includegraphics[width=.8\columnwidth]{Fig.s/Error.png}
% \caption{Qiskit~\cite{Qiskit} size error raised on a BVn-14 benchmark~\cite{li2020qasmbench} circuit on a 5 qubits Fakelima~\cite{FakeProvider} system}
% \vspace{-3mm}
% \label{fig:Error}
% \end{Fig.}

%%%%%

%%%%%

\subsection{Related Work}\label{sec:related}
In this part, we discuss quantum compilation methods,  quantum programming languages, and most recent  accomplishments in this area of research. While quantum computing is still in its infancy (in terms of both hardware or software), its potential advantages over so-called classical computing for particular algorithms, e.g., in the context of drug discovery, machine learning and prime factorization, are very promising~\cite{kandala2017hardware,shor1999polynomial,peruzzo2014variational}.  Quantum  systems are currently being heavily researched, and the major efforts focus on the areas of  compiler  support~\cite{li2022paulihedral,litteken2020updated,shi2020resource,nguyen2022retargetable,booth2012quantum},  operating system support~\cite{ravi2021adaptive}, and programming languages~\cite{javadiabhari2015scaffcc}.      
   
In particular, several quantum programming languages,  including Q\#~\cite{svore2018q}, OpenQASM  3.0~\cite{cross2021openqasm}, Silq~\cite{bichsel2020silq}, Scaffold~\cite{javadiabhari2015scaffcc}, Scaffcc~\cite{javadiabhari2015scaffcc}, and QCOR~\cite{mccaskey2021extending,mintz2020qcor}, have been developed over last couple of decades. It is to be noted however that, at present time, these languages are very close to the low-level assembly code and depend partly on user-supplied gate insertions. For example, Qiskit (IBM's open-source software development kit for working with quantum computers at the level of circuits, pulses, and  algorithms~\cite{Qiskit}) employs such a strategy.  

Compilation of quantum programs consists of no more than three main {\em stages}: i) matrix-to-gates conversion, ii) IR optimizations, and iii) logical-to-physical qubit mapping and circuit execution. In this context, a {\em matrix} represents a function/system which is applied on sample inputs (a vector of inputs) to generate a final output (an output vector). In the first step, the matrix is translated into 1-qubit and 2-qubits gates using an algorithm such as Fowler~\cite{fowler2011constructing,fowler2012surface}. If, on the other hand, the programmer encodes the quantum circuit directly (i.e., if he/she inputs the gates instead of the matrix), then the preceding phase -- Fowler gate production -- can be omitted. The Fowler~\cite{fowler2011constructing} method is probably the most well-known (first-stage) compilation technique, which takes a target matrix of the Hilbert space as input and searches for gate matrices at each step in order to {\em approach} the target matrix within a certain proximity, called the Fowler Distance, which is calculated as follows:  
\begin{center}
$dist(U,U_l)\;=\;\sqrt{\frac{2\;-\;tr\;|U\;.\;U_l^\dag|}{2}}. $ 
\end{center} 
Note that Fowler Distance is calculated over multiple iteration and at each step, this distance is calculated until we reach the desired threshold. Unfortunately, the complexity of this  approach can be exponential; hence, it is preferable for a quantum programmer to input the circuit using quantum gates or input a circuit-matrix hybrid, which consists of different circuit parts; each has either a gate representation or a matrix representation. Therefore, the complexity of the Fowler algorithm can be reduced, thereby improving its efficiency. Current compilers try to optimize this procedure by reducing the exponent base by restricting the collection of existing gates. Booth et al.~\cite{booth2012quantum} offer further optimizations by i) using a bidirectional search and ii) modifying the basis to a modified Pauli basis to facilitate the Fowler distance computations. 

The second stage of a quantum compiler performs IR (intermediate representation) level optimizations. Note that the IR abstraction layers employed by compilers targeting quantum programs range from source-code level abstractions to system-level abstractions. In modern quantum compilers, LLVM~\cite{llvm} is the standard, which is close to machine-level IR. Recent studies have implemented a variety of LLVM-based optimizations aimed at various domains~\cite{nguyen2022retargetable}. For example,  Paulihedral~\cite{li2022paulihedral} is one of the most recent works; it proposes retaining a gate matrix IR abstraction until the final levels for more straightforward circuit optimizations, such as depth reduction, gate cancellation optimization and swap reduction, by ignoring false dependencies (i.e, dependencies that are in the DAG representation of the circuit due to the order of operations, while the order is not actually important) between layers of gate production and re-ordering the gate operations. This approach fits well to enhance our work since discovering false dependencies at the final stage of compilation (where our approach is embedded) would be complex. It can also be used to arrange gate layers according to the volume of their true dependencies in order to maximize serialization opportunities. 

The last stage of the compilation of a quantum program involves quantum circuit-level optimizations including the {\em mapping} of logical qubits to physical qubits, gate cancellation, swap reduction, etc. Optimizations at this stage focus mostly on boosting the reliability of the output. Works such as~\cite{du2001implementation,ohkuracrosstalk,liu2021qucloud} concentrate on the important problem of {\em mapping} logical qubits (qubits in the quantum program) to physical qubits (qubits that are physically implemented in the target quantum machine) to improve reliability. On the other hand, Murali et.al.~\cite{murali2020software} study gate scheduling to boost reliability. Other efforts focus on optimizing the number of shots~\cite{arrasmith2020operator}, dynamic decoupling optimizations for reliability~\cite{10.1145/3466752.3480059}, ancillary reuse~\cite{parent2015reversible,parent2015reversible}, and system selection optimizations~\cite{ravi2021adaptive}. Our approach proposed in this paper operates at this last stage compilation of a quantum program. Note  however that we differ from the related work in that our approach i) focuses on minimizing the number of qubits to prevent  "system size compilation errors"; to offer the possibility of getting output from a large quantum circuit considering system  size limitations; and ii) increasing the circuit's output reliability via serial execution. 

%%%%%

%%%%% 

\section{Motivation and Problem Definition}\label{sec:motiv} 
This section describes the three main factors that have motivated us to design a compiler-based strategy for minimizing the size of a given quantum circuit. We have identified two major impediments to raising the qubit count on existing NISQ systems. Additionally, we noticed that some of the well-known quantum circuits, such as Bernstein-Vazirani~\cite{bernstein1997quantum,du2001implementation,middleresetIBM}, can be converted into more "serialized"  circuits by using the MM and MR gates in the IBM QAOA systems. 

\subsection{Size Limitation in NISQ Systems}
The continuous demand for more processing power requires the development of quantum computers with large number of physical qubits. While the addition of a single qubit may result in an exponential rise in the compute capability of a quantum system, the absence of a physical "quantum memory" places the whole processing weight on qubits. While QAOA systems combine quantum computing with a classical memory, the quantum states still remain "unmemorable". 

With all these demands for more qubits on quantum computers, the reliability of physical qubits and their connecting links remain as the main issue. Furthermore, as the system grows in size, the reliability degradation becomes even more difficult to avoid. This results in a decrease in the number of links between qubits to eliminate crosstalk noise~\cite{ash2020experimental}, and as a result, leads to more scattered/distributed qubits, lowering the qubit processing speeds. Nonetheless, owing to the exponential expansion of processing power, this cost of processing speed may not be the primary concern. Still the {\em decrease in the number of links in larger NISQ systems}, needs the inclusion of multiple extra SWAP operations to execute the quantum circuit. 

\subsection{Reliability Concerns with Larger Systems}
The majority of qubits in current IBMQ system architectures, for example, have two links, whereas a few have one or three links. A large quantum circuit will have various "entangled"\footnote{Entanglement is a technique used by quantum computers to create a "correlated state" across  multiple qubits, where changing one affects the other.} operations, necessitating the use of numerous SWAP gates to complete the execution. This not only results in extra gate errors but also in excessive crosstalk and coherence faults;  the latter is owing to possible differences in the completion times of different qubits~\cite{Liu2022not}. Thus, the recently-proposed logical-to-physical qubit mapping techniques, e.g., NASSC~\cite{Liu2022not} and Qcloud~\cite{liu2021qucloud}, try to reduce such SWAP costs. In particular, recently, quantum computing research has primarily focused on mapping~\cite{liu2021qucloud},  scheduling~\cite{murali2020software,wille2013improving,ohkuracrosstalk}, and even architectural-level strategies~\cite{xie2022suppressing}, for reducing crosstalk errors. On the other hand, to minimize coherence faults, dynamic decoupling strategies have been developed, the most recent of which is ADAPT~\cite{10.1145/3466752.3480059}.  

While these strategies can help us reduce various errors in quantum circuits, errors cannot be completely eliminated,  owing to the physical properties of qubits and quantum gates.
For example, executing a 12-qubit circuit like  Bernstein-Vazirani, which can be categorized as one of the relatively simple "medium-size" quantum circuits, on a sample IBMQ device results in entirely unreliable outputs (a fidelity of 0.007 is reported in~\cite{middleresetIBM}). However, by employing the circuit minimization technique, this fidelity can increase up to 0.31 (400x faster compared to the prior case~\cite{middleresetIBM}).  Consequently, there is a strong motivation for exploring strategies that minimize the qubit requirements of a given quantum circuit in the absence of a quantum memory, and to our knowledge, not only there is no published algorithm that can do this minimization for general circuits, but also there is no solution for the MM delay problem, which will be detailed  later in the paper.    

\subsection{How to Resize the Circuit via the MM and MR Gates?} 
The first step in building a circuit for the current NISQ machines is defining the required number of qubits and classical registers for measurement. The programmer encounters an error if the number of qubits in the circuit to be implemented exceeds the system size. For instance, defining a Bernstein-Vazirani circuit with a size of 10 logical qubits and attempting to execute it on a system with 7 physical qubits results in a compilation error. In reality, since current quantum systems support both the MM and MR gates, qubits can be {\em reused} to incarnate (and execute) additional qubits, and thus this error can be resolved. As stated in \cite{middleresetIBM}, the Bernstein-Vazirani circuit of any size can be executed, in principle, {\em sequentially} using no more than 2 qubits via the MM and MR gates. While executing this circuit sequentially contradicts the algorithm's stated goal of demonstrating the parallelism intrinsic in quantum computing, it makes sense on a NISQ system. This is because, as previously mentioned, the NISQ systems depend on SWAP gate operations, which results in the circuit running  sequentially. While this potential has been highlighted before~\cite{middleresetIBM}, to our knowledge, no works have been offered to use it and prevent this error. 

\begin{figure}[t!]
\center
\includegraphics[width=\columnwidth]{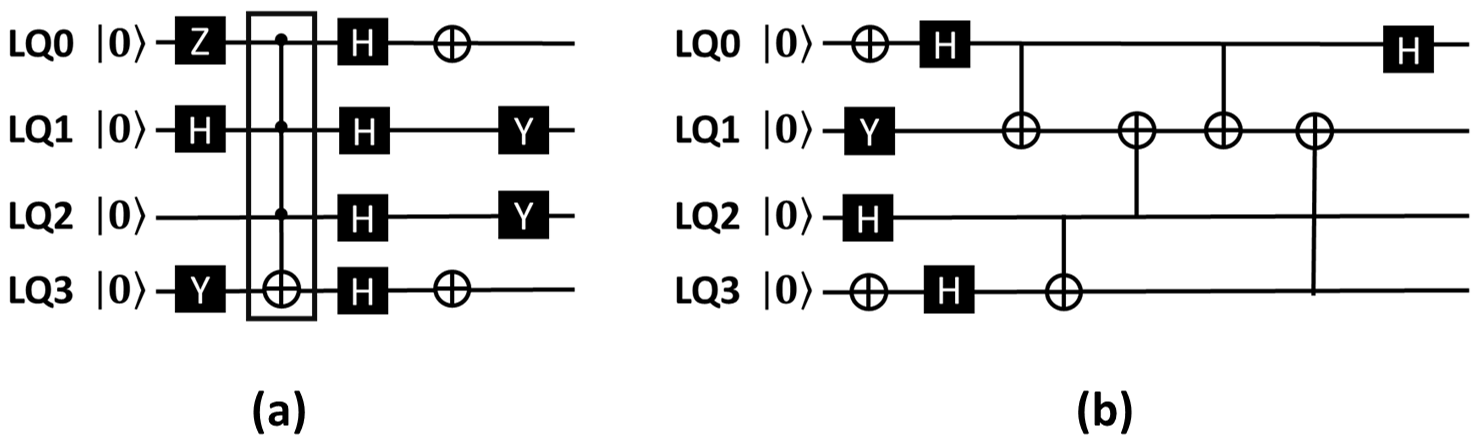}
\vspace{-3mm}
\caption{Two unresizable circuits.}
\vspace{-3mm}
\label{fig:Unresizable}
\end{figure}

\begin{figure}[t!]
\center
\includegraphics[width=.5\columnwidth]{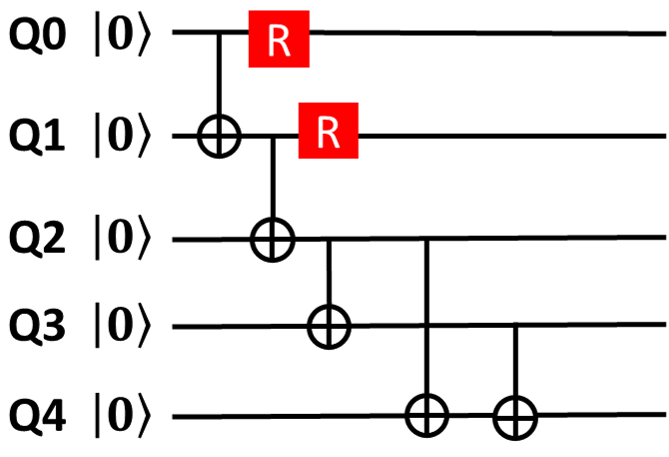}
\caption{A resizable circuit.}
\vspace{-5mm}
\label{fig:resetable}
\end{figure}

Although, as illustrated in Fig.~\ref{fig:Unresizable}a, not only a fully-entangled circuit --like Fig.~\ref{fig:Unresizable}a-- cannot be reduced to a smaller circuit, but also it requires extra ancilla qubits to generate this large gate from the physical available gates. These circuits are primarily constructed using compiler techniques based entirely on quantum theory and make assumptions that are more compatible with a theoretical quantum computer. In comparison, current compilers like Qiskit~\cite{Qiskit} and SCAFFC~\cite{javadiabhari2015scaffcc} are more considerate of the circuit design constraints imposed by existing hardware. %This can also be advantageous for them in terms of compilation time complexity.  
For instance, Paulihedral~\cite{li2022paulihedral}, a state-of-the-art compiler-based approach, generates quantum circuits using only unitary and CNOT gates, which have a higher chance of producing resizable circuits. These circuits can achieve significant size reductions when certain constraints are met, which will be elaborated later in the design section. 
\textbf{A circuit is "resizable" if it contains at least one qubit that may attain its final state without the need of all other qubits.}
Fig.~\ref{fig:resetable} depicts an example quantum circuit in which $q_0$ and $q_1$ complete their tasks in this fashion. For $q_0$ to finish its task, we only need $q_0$ and $q_1$, and for $q_1$ to finish its task only $q_0$, $q_1$ and $q_2$ are required. Hence, the constraint stated above is satisfied, and this circuit is resizable.

Our goal in this paper is to minimize the size of a given quantum circuit by running it in a serial/sequential manner by employing the MM and MR gates, and also to optimize output reliability for circuits executing in architectures with limited number of links per qubit. Thus, our main novelties include i) giving a polynomial time algorithm to minimize the size of a given quantum circuit; ii) providing an implementation of this algorithm and proving that it really minimizes the input circuit; and iii) avoiding the reliability issues due to delays of MM gates, which happen through multiple shots.

%%%%%

%%%%%
\vspace{-3mm}
\section{Overall Design}\label{sec:design} 
The first step to shrink the size of the circuit is to identify the criteria which makes a circuit serially executable. Based on our observations, the only criteria is:

\begin{quote}
\textbf{A circuit is "serializable" if and only if there exists a qubit that can complete its final gate operation without the activation of other qubits on the circuit.} 
\end{quote} 

\begin{figure}[ht!]
\center
\includegraphics[width=.6\columnwidth]{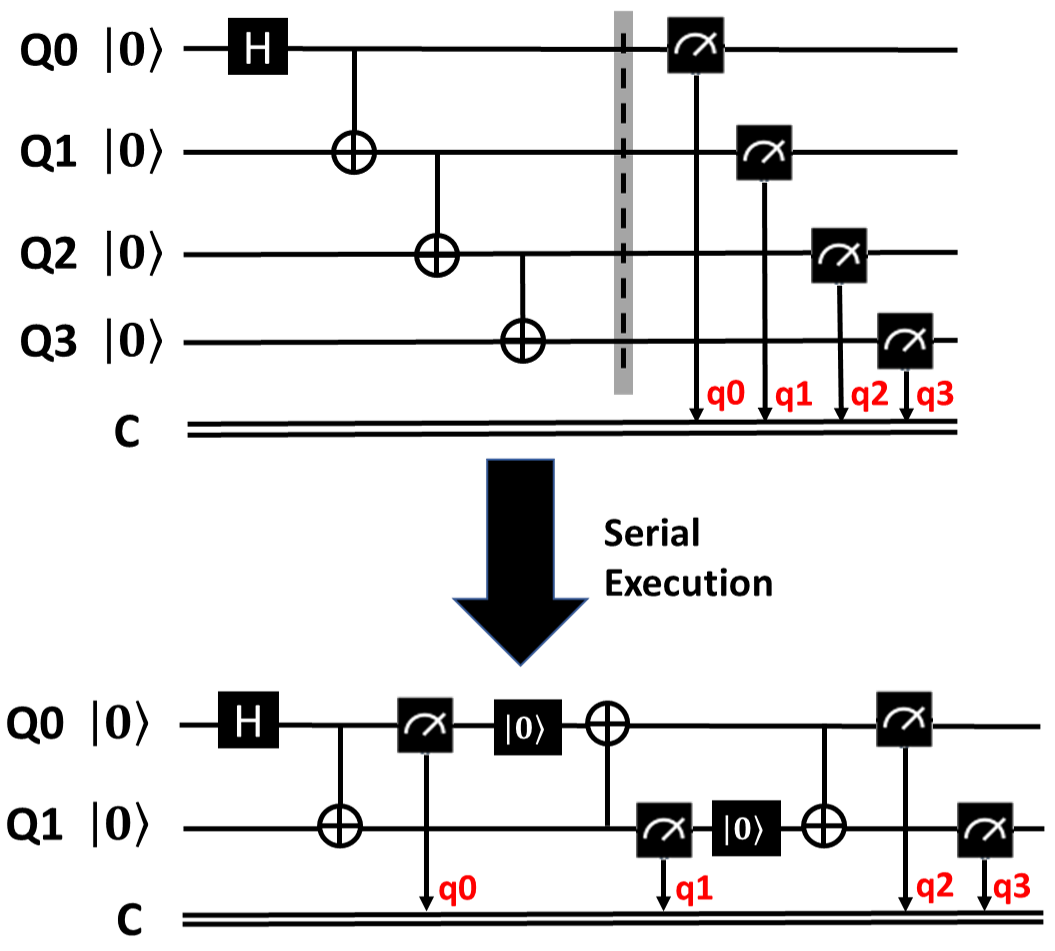}
\caption{A sample cat\_state\_n4 circuit and its serial execution.}
\vspace{-6mm}
\label{fig:Serial}
\end{figure}

Fig.~\ref{fig:Serial} shows a serializable circuit and its serial execution with the minimum necessary qubits, whereas Fig.~\ref{fig:Unresizable} depicts two circuits that do not meet our serializability requirement. As an example, in Fig.~\ref{fig:Unresizable}A, none of the qubits can complete its operation without activating the remaining qubits. On the other hand, in the circuit shown in Fig~\ref{fig:Serial}, $q_0$ can complete its task with the assistance of only $q_1$. Therefore, the first circuit should be executed in parallel, while the second circuit can be serialized (i.e., its size can be reduced).

\subsection{Requirements for Circuit Size Reduction (Sequential Execution)} 

In this section we carefully explain the reasons why the circuits shown in Fig.~\ref{fig:Unresizable}A and~\ref{fig:Unresizable}B are not resizable (cannot run serially). First, let us discuss the use-cases for which the MM/MR gates are beneficial. The MM gate enables users to {\em measure} a qubit {\em after} it has completed its final action, while the MR gate is utilized to {\em reset} it to a $|0\rangle$ state. Together, these two operations enable any result qubits to be measured and reset so that it can be used by other qubits. For garbage qubits\footnote{Garbage qubits are qubits that are employed to help to generate the output of the program but are not themselves output of the circuit.}, on the other hand, only an MR gate is needed (for them to be resued).

Now let us discuss the reason behind why Fig.~\ref{fig:Unresizable}A and Fig.~\ref{fig:Unresizable}B are not resizable. For Fig.~\ref{fig:Unresizable}A, we can see that none of the  qubits this circuit possesses can perform its task/operation in isolation from the other qubits owing to the gate that {\em entangles} them all (i.e., complete entanglement). Therefore, resetting any of these qubits is not beneficial for shrinking the size of the circuit.
Fig.~\ref{fig:Unresizable}B shows a more complex circuit to demonstrate the criteria that should be satisfied in order to serially execute a circuit as much as possible. In this example, Q0 is used in OP1 and OP4, needing Q1 for its completion. However, OP2 and OP3 should also be executed before we can reuse Q0 (OP2 and OP2 should finish before OP4), meaning that Q0 needs all the qubits to be available when it finishes its last operation. Q1 has operations with all the other qubits, eliminating the possibility that it can be reused for serial execution. Q2 has operations with Q1 and Q3 (OP3 and OP2), still, it needs OP1 to be executed beforehand, meaning it needs all the qubits to be available before its final operation. Q3 is similar to Q2 since it does not have any operation with Q1, still needing OP1 to be executed first. Therefore, we cannot reuse any of these qubits, eliminating the possibility that they can be used as a choice for resizing the circuit. Therefore, even if MM and/or MR gates are employed on the qubits, it does not provide any benefits since no qubits exist to be executed on the reset qubit. Thereby, it denotes this circuit is not resizable and cannot benefit from our scheme. Based on these examples, we introduce the concept of {\em activation}, meaning that a qubit can finish its task without directly or indirectly using all other qubits. Note that not having an operation with other qubits does not satisfy the activation requirement since there are cases where qubits do not have an operation with each other but still need each other to complete their operations first, such as Fig.~\ref{fig:Unresizable}B. It is important to note that the order of operations is extracted from DAG, which may contain false dependencies. In our scheme, we do not differentiate between false and true dependencies since we do not want to change the structure of the circuit. Still, it is possible to feed the algorithm with the true dependencies and reorder the operations if possible, but our approach mainly focuses on changing the execution and not the circuit.

The most beneficial qubits for resizing/serializing a given quantum circuit are those qubits that can complete their tasks with the minimum number of activation from other qubits. By prioritizing the qubits based on the number of activation from other qubits, in our algorithm, we reduce the circuit size as much as possible. Therefore, in our algorithm, we introduce an additional constraint aiming at maximizing the improvement (note that this is {\em not} a   constraint for resizability of the circuit; rather, it is a criterion to maximize the potential improvements):
\begin{quote}
\textbf{The best nominee for resizing a circuit is the qubit that has the lowest number of other qubits that need to be activated.} 
\end{quote} 

One way of checking for these attributes is to use a {\em circuit DAG}, as a DAG precisely captures the sequence of operations, dependencies, and the stages at which each qubit completes its task. The only aspect that DAG does not capture is false dependencies, which are not important for us due to three reasons:
\begin{itemize}
    \item Finding false dependencies is exponentially complex for a given circuit and has polynomial complexity in the matrix production stage.
    \item False dependencies can only cause an adverse effect in our circuit minimization algorithm in one case, which is not expected to be frequent. We further elaborate on this in Section~\ref{sec:proof}. 
    \item Our main goal is to minimize the circuit only through changes in the execution strategy, not through the changes in the circuit itself or its gate operation order.
\end{itemize}
In any case, false dependencies can be addressed in the IR-level optimization stage efficiently, as demonstrated by~\cite{li2022paulihedral}, and our algorithm can perform both with true dependencies and DAG dependencies as inputs without changing the algorithm itself. Providing an efficient algorithm to obtain true dependencies as an input to the algorithm is left for future research and is out of the scope of this article.

%%%%%%%%%%%%%%%%%%%%%%%%%%%%%%%%%%%%%%%%%%%%%%%%%%%%%%%%%%%%%%%%%%%%%%%%%%%%%%%%%%%%%%%%%%%%%%%%%%

\subsection{Our Proposed Algorithm for Circuit Minimization}\label{sec:alg}
\begin{figure*}[t!]
\center
\vspace{-4mm}
\includegraphics[width=.55\textwidth]{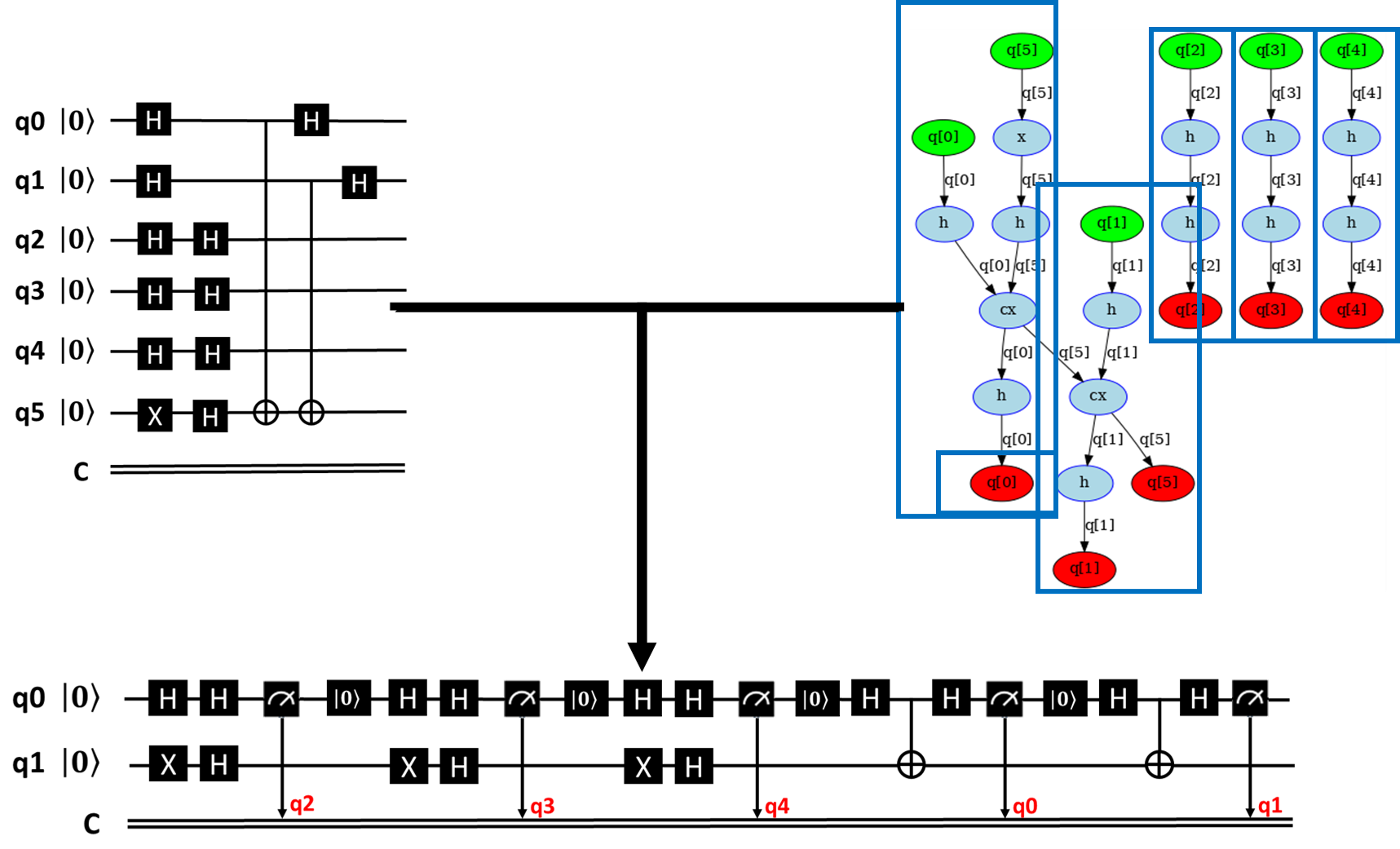}
\caption{A 5-qubit Bernstein-Vazirani circuit and its DAG representation, depicting the steps of our proposed algorithm for circuit size minimization via serial execution.}
\vspace{-6mm}
\label{fig:BVN}
\end{figure*}

This section introduces our approach and goes through an example scenario step-by-step to demonstrate how it works in practice. Please keep in mind that we are attempting to reduce the size of the final circuit generated by the compiler and/or programmer. Any optimizations at the gate level or for handling the false dependencies on the circuit at an earlier stage would change the circuit gates and their order and should be applied before invoking our proposed approach. In fact, such techniques would help our approach to further enhance its circuit minimization potential. For instance, circuits constructed entirely of unitary gates and CNOT gates will be more favorable to our proposed approach, as can be seen in Fig.~\ref{fig:Serial}. Similarly, reordering/repositioning gates by ignoring false dependencies would aid in reducing wait-times as well as coherence errors, and may aid with further serialization (they should typically be performed by the intermediate stage of the compiler, since they are easier to detect and handle at that level). 

Algorithm~\ref{alg} gives our proposed algorithm for quantum circuit resizing. To begin, we look for a qubit that can complete its task with the fewest number of other qubits being activated. For instance, if a qubit consists entirely of unitary gates, it would perform its function without activating any other qubits.
We obtain this information by tracing back the DAG for each qubit from {\em leaf} to {\em root}.
We add the gate operations required before the chosen qubit completes its task, keeping track of the total number of gate operations added in each qubit line. Then, we deduct this value from the total number of gate  operations performed in each of the activated qubits by the added gates.
This is done mainly to avoid repeating the gate operations to maintain the correct circuit.
Note that, when this value reaches zero, we perform the MM/MR procedure and we will look up for the next logical qubit to load into the physical qubit that has been reset, for sequential execution.
We update the dependency lists by deleting the qubits that have already been allocated from the current list and then search for the next addition.
This ensures that we prioritize serialization by attempting to increase the number of added qubits in each iteration as little as possible. 
\RestyleAlgo{ruled}
\begin{algorithm}
\scriptsize
\caption{Our proposed resizing algorithm.}\label{alg}
\textbf{Input:}
\BlankLine
DAG circuit
\BlankLine
\textbf{Register:}
\BlankLine
$D-list[q]\longleftarrow$ List of dependencies for $q$
\BlankLine
$L-list[D]\longleftarrow$ list of $D-list$s
\BlankLine
$Count[q] \longleftarrow List\, of\, number\, of\, q's\, gate\, operations$
\BlankLine
$PQ \longrightarrow$ A flag list to show available physical qubits
% \For{q \textbf{in} circuit}{
% $count[q] \longleftarrow number\, of\, qubit's\, gate\, operations$
% \BlankLine
% }
\BlankLine
\textbf{Output:}
\BlankLine
$New-circuit$
\BlankLine
\For{qubit \textbf{in} DAG}{
$D-list[qubit] \longleftarrow [qubit]$
\BlankLine
{Trace back in DAG from qubit last level to its root and append all qubits it interacted with to the list}
\BlankLine
$L-list.append(D-list)$
% \BlankLine
% $L-list.sort(key=lambda\;i:len(i))$ \tcp*{sort this list of D-lists by list's length}
}
\BlankLine
$L-list.sort(key=lambda\;i:len(i))$ \tcp*{sort this list of D-lists by list's length}
\BlankLine
% \SetKwFunction{FSU}{SelectandUpdate}
% \FSU{$PQ$}{\{
%         \BlankLine
%         $Chosen\longleftarrow L-list[0].pop()$
%         \BlankLine
%         \KwRet L-list
%         \}
%     }
% \BlankLine    
$New-circuit \longleftarrow \phi$
\BlankLine
$PQ\longleftarrow$ Is a list of Size L-list first element, initialize to $0$ meaning available
\BlankLine
\textbf{Algorithm:}
\BlankLine
\eIf{len(L-list[0]) == circuit's number of qubits)}{
\textbf{print}(circuit is not resizable)
\BlankLine
$New-Circuit = circuit$
\BlankLine
\textbf{break}
}{
\While{(len(L-list) != 0)}{
$Chosen\longleftarrow L-list.pop(0)$
\BlankLine
$l-list\,=\,[[q\textbf{ for }q\textbf{ in }D-list\textbf{ if }q\textbf{ not in }Chosen]\textbf{ for }D-list\textbf{ in }L-list]$
\BlankLine
$L-list.sort(key=lambda\;i:len(i))$
\BlankLine
\tcp*{Update the D-list elements of L-list based on the qubits that became activated and resort the L-list to ensure minimum addition -if any at all- of qubits will happen at each iteration}
Assign logical qubit from $Chosen$ to physical qubits available in $PQ$, add elements to PQ if $Chosen$ can't fit in available
\BlankLine
Update new $PQ$ values to $1$ meaning occupied
\BlankLine
\For{qubit \textbf{in} Chosen}{
$New-Circuit \longleftarrow New-Circuit\,+$ Add gate operations if gates are in $Chosen$ --only--
\BlankLine
Update count values in $Count[q]$ list 
\BlankLine
\If {$Count[q]==0$}{
$PQ[q]\longleftarrow0$
}
}
% {New-Circuit: append every gate operation of Chosen and update the counts values after gate additions in count until all possible count elements become 0 - using only the qubits in chosen- and keep a mapping of new names to old names. MM -if applied- must happen to the same classical register as the original circuit, MR them and set them as free logical qubits for the next iterations.}
% \BlankLine
% $l-list\;=\;[[q\;\textbf{for}\;q\;\textbf{in}\;D-list\;\textbf{if}\;q\;\textbf{not in}\;Chosen]\textbf{for}\;D-list\;\textbf{in}\;L-list]$
% \BlankLine
% $L-list.sort(key=lambda\;i:len(i))$
% \BlankLine
% \tcp*{Update the D-list elements of L-list based on the qubits that became activated and resort the L-list to ensure minimum addition -if any at all- of qubits will happen at each iteration}
}
}

\end{algorithm}

Fig.~\ref{fig:BVN} shows an example application of our proposed algorithm on an example Bernstein-Vazirani circuit. In this example, $D-list$ for $q_0$ is $[q_0,q_5]$; and for $q_1$, it is $[q_1,q_5,q_0]$ where  $q_0$ is a false dependency, meaning this is an order of operation in the DAG, and these two operations can change their ordering. On the other hand, $D-list$ for $q_2$ is  $[q_2]$; for $q_3$, it is $[q_3]$; for $q_4$, it is $[q_4]$; and finally, for $q_5$, $D-list$ is $[q_5,q_1,q_0]$.
Now, we have the sorted $l-list\;=\;[[q_2],[q_3],[q_4],[q_0,q_5],[q_1,q_5,q_0],[q_5,q_1,q_0]]$ as described, the first element will be $[q_2]$, we will assign all the qubits in the dependency list to the first available physical qubits, which in this case means only assigning $q_2$ to first available physical qubit (Q0\footnote{Uppercase Q is for physical qubits}).
Then, we will remove all the assigned qubits from all the elements in $l-list$, which means removing $q_2$ from all the elements of $l-list$. Additionally, we also update the count list, which shows the number of unexecuted operations for each qubit. In this case, all the operations for $q_2$ are already executed ($count[q_2]=0$), then we will perform the MM and MR on the physical qubit assigned to $q_2$ and look for a new qubit for assignment. The new updated list is $l-list\;=\;[[q_3],[q_4],[q_0,q_5],[q_1,q_5,q_0],[q_5,q_1,q_0]]$. We perform the same steps for $q_3$ and $q_4$, assigning them to the same physical qubit (Q0).
After these two step, $L-list$ becomes $l-list\;=\;[[q_0,q_5],[q_1,q_5,q_0],[q_5,q_1,q_0]]$.
In the next step, we choose $[q_0,q_5]$ which is the element for $q_0$. Since it contains more than one qubit, we assign the first logical qubit ($q_0$) to the first available physical qubit (Q0) and assign the second logical qubit ($q_5$) to the second available physical qubit (Q1).
By scheduling the gates for these two qubits and updating the list accordingly, the $count[q0]$ will become 0, meaning we executed all of its gates. In the next step, we will apply MM and MR on Q0 and look for a new logical qubit for the assignment. After the update, the new $L-list$ will be $l-list\;=\;[[q_1],[q_1]]$.
We choose the first element, $[q_1]$, and assign $q_1$ to the first available physical qubit (Q0). After scheduling all the operations and updating the list, we see that both $count[q_1]$ and $count[q_5]$ are equal to 0. It means that all the operations for both $q_1$ and $q_5$ are executed, and we can measure and reset them using MM and MR gates if needed. Since $q_5$ is garbage, it does not have any measurement operation. Finally, we will empty the $l-list$, and the algorithm is done.

%As described in Algorithm~\ref{alg}, we next pop  $[q_0,q_5]$ as the Chosen (the new logical qubit to be added to the physical qubit that has been reset); the first element is the representative of the qubit that will be MM/MR, meaning that its gate operation count should go all the way to $0$. We write down all the gate operations before $q_0$ as the first element finishes its job; the gate count in $q_0$  becomes 0 and MM/MR adds to it with the same logic as the previous ones, and the gate count on $q_5$ becomes  $count[q_5]\;=\;4\;-\;3\;=\;1$, meaning that we cannot free $q_5$ yet. $q_0$ and $q_5$ occupy logical $q_0$ and $q_1$ respectively; the logical $q_0$ is released and the new $L-list$ will be $l-list\;=\;[[q_1],[q_1]]$. We pop off $[q_1]$ and load its operation into the released logical $q_0$, the counts for both $q_1$ and $[q_5]$ become 0; so,  we perform MM/MR on logical $q_0$ with the attributed classical register to $q_1$ and we only perform MR on $q-5$ since there exist no measurement assigned to it, and the new $L-list$ becomes $l-list\;=\;[[q_1]]$.  
%We pop off $[q_1]$ as the Chosen again but this time the count is already 0; as a result, no further action is needed and  $L-list$ finally becomes $l-list\;=\;\phi$.  

% As shown in the example, the false dependency only causes an extra null iteration and does not affect the logic of the algorithm.
Investigating the impact of reordering the false dependencies at the IR optimization stage, e.g., for reliability optimization purposes, is beyond the scope of this study and  left to a future study. Also, producing a circuit considering false dependency reordering can aid in further circuit reduction; however, how to optimize for it is also left for future exercise. As of now, the only possible case where false dependency can lead to a higher qubit count is when the qubit selected in an iteration has the fewest true dependencies. Still, if the qubit has multiple false dependencies, it will not be chosen for execution (since the total dependencies are considered) and can lead to serializing the circuit in another fashion. Therefore, it may cause our circuit not to be entirely serialized. We want to mention that this scenario has not happened in any of the circuits we tested in this study. If it happens in some other circuits, however, we want to emphasize that it does not affect the final output of the circuit; rather it will possibly cause the circuit to be minimized by a smaller degree. It is also to be noted that changing the circuit to order it based on true dependencies will change the whole circuit implementation~\cite{li2022paulihedral}. 

\subsection{Proof}\label{sec:proof}
As previously noted, the presence of false dependencies may or may not affect the circuit minimization, but detecting them at the last compilation stage is inefficient. Instead, the easiest place to look for them is probably the IR optimization stage, as in  Paulihedral~\cite{li2022paulihedral}, where the circuit construction takes place at a matrix-multiplication level. Therefore,  optimizing the sequence of operations for minimization -- if they have any influence -- would be a more straightforward process in those stages and would probably lead to a totally new circuit. As far as this work   is concerned, we would like to achieve minimization via changes only in circuit execution, not in the design of the circuit itself. We save the study of designing "serialization-friendly" circuits for a future study.  

We now present a formal proof showing that our proposed approach {\em  minimizes} the given quantum circuit, assuming that all dependencies are true dependencies (algorithm input can be a true dependency list; we obtain dependencies from its DAG) or at least the operations leading to false dependencies have already been ordered in the best manner for size reduction (please note the dependencies, in this context, are  algorithm inputs and not part of the algorithm itself). We use proof by contradiction (\textbf{this is a prove of minimization due to a change in execution not a change in circuit}):
\textbf{Please note that any addition of qubits at any stage or iteration will permanently increase circuit size unless loaded on a reset qubit.}
Suppose that $q_i$ is a qubit with more dependencies than minimum dependency $q_m$ at iteration $k$; so, it can lead to a smaller circuit than $q_m$ ($S[q_x]$ represents $q_x$ dependency list size). There are three different scenarios to consider: 
\squishlist
    \item The $q_m$ dependency list is a subset of the $q_i$ dependency list. Obviously, in the dependency list of $q_m$, the first element is $q_m$ itself (it will be in the $q_i$ dependency list as well). Note that adding $q_i$ will increase the circuit size by $S[q_i] > S[q_m]$ qubits permanently. Further, adding $q_m$ even right before $q_i$ will lead to $S[q_i]\;=\;S[q_i]\;-\;S[q_m]$ and release $q_m$ for $q_i$ for use, thereby increasing the size of the circuit at most by    $S[q_m]\;+\;S[q_i]\;-\;S[q_m]\;-\;1$, leading to a   smaller circuit, which is clearly a contradiction.    
    \item The $q_i$ and $q_m$ dependency lists have no intersection. In this case, they are likely to be two separate circuits ($C_1$ and $C2$) we are trying to resize (in the example Bernstein-Vazirani circuit above, based on the dependency lists, for our purposes, we can consider $[q_0,q_1,q_5],[q_2],[q_3],[q_4]$ as separate circuits),  with $C_1$ being the circuit containing $q_i$ and $C_2$ being the one that has $q_m$. Since they are like two separate circuits, it is clear that any minimization needs them to work in a serial fashion, meaning that either we start with $C_1$ and then load $C_2$ or vice-versa. In either case, irrespective of the selection order of $q_m$ or $q_i$, the final minimum-sized circuit will be of size $\textbf{Max}\{C_1,C_2\}$, leading to the two cases with an equal number of physical qubits used and will be a contradiction.
    \item The $q_i$ and $q_m$ dependency lists intersect, but the $q_m$ dependency list is not a proper subset of the $q_i$ dependency list. In this scenario, we have one of the following two cases: 
    \squishlist
        \item $q_m$ is an element of the $q_i$ dependency list, which is similar to the first case with $q_m$ dependency list is a subset of $q_i$ dependency list.
        \item $q_m$ is not an element of the $q_i$ dependency list.  In this case, $q_m$ adds fewer  qubits than $q_i$ but guarantees at least one reset qubit, which provides one available qubit for $q_i$. It is worth to note that, due to the intersection of $q_m$ and $q_i$ dependency lists, some elements of $q_i$ dependency list would be already added to the target circuit. Now, by adding the remaining elements of $q_i$, a target circuit with the size of $q_i$ dependency list is created, given that $q_m$ is already in the target circuit. Hence, selecting $q_m$ leads to a final circuit with same or less number of qubits compared to selecting $q_i$ first, which is contradicts our assumption. 
    \squishend
\squishend

Therefore, by using contradiction, we prove that our algorithm can minimize the size requirement if changing the order of operations is not attainable.
%%%%%%%%%%%%
\subsection{Complexity Analysis of the Proposed Algorithm}\label{sec:complexity}
%%%%%%%%%%%%%%%
In this section, we study the timing complexity of our proposal. First, the algorithm must extract dependencies from the DAG of the circuit. As indicated in  Section~\ref{sec:background}, the roots and leaves of the DAG represent qubits, the remaining nodes represent gates, and the edges capture the qubits of the corresponding gate operations. Assuming a circuit with $n$ qubits and $m$ total gate operations, we need $n$ qubits to check from leaf to root and at most $m$ operations to check for each qubit;  hence, this phase of our algorithm takes $\mathcal{O}(nm)$  complete.     

The subsequent step of the algorithm consists of circuit   resizing based on the dependency lists of the qubits obtain  in the previous phase and saved as a list of lists named  l-list.  The outside loop is a while-loop on all $n$  dependency lists of qubits, and inside we sort this l-list based on the dependency list size (each dependency list is an element of l-list), which accounts for $\mathcal{O}(n\log{}n)$ and may be optimized to $\mathcal{O}n$ if implemented by locating the minimum size element of l-list at each iteration. Note that each  iteration includes the addition of a maximum of $m$ gates; therefore, this phase of the algorithm takes at most $\mathcal{O}(mn^2\log{}n)$. Overall, the complexity of the algorithm is: $\mathcal{O}(nm)+\mathcal{O}(mn^2\log{}n)=\mathcal{O}(mn^2\log{}n)$.

Based on the results, we want to emphasize that our approach to optimizing (serialize) a given circuit has a polynomial complexity, which is good. For example, trying to resize a circuit of size 1000, with one million gates to a minimum 100 qubits takes less than 10 seconds on a 2016 Intel core i7 6950X system.

%on a 2016 Intel core i7 6950X 320,440 MIPS~\cite{Wiki} speed at 3.5 GHZ will take less than 10 seconds.

\begin{table*}[ht!]
    %\vspace{-4mm}
    \centering
    \begin{tabular}{ |p{3.2cm}||p{2.5cm}|p{2.5cm}|p{2.5cm}|p{2.5cm}|p{2.5cm}|}
        \hline
        \multicolumn{6}{|c|}{PST report on ibmq-lima} \\
         \hline
         Circuit Name & \# of Qubits in Normal Execution & \# of Qubits in Sequential Execution & PST & Total Gate Count & CNOT Gate Count \\
        \hline
        bv\_n14~\cite{cross2021openqasm}   & 14    & 2 &  51.2\% & 121 & 13 \\
        bv\_n19~\cite{cross2021openqasm} &   19  & 2   & 42.2\% & 166 & 18 \\
        wstate\_n27~\cite{cross2021openqasm} & 27 & 3 &  35.4\% & 593 & 124 \\
        ghz\_state\_n23~\cite{cross2021openqasm}    & 23 & 2 & 52.2\%  & 70 & 22 \\
        swap\_test\_n25~\cite{cross2021openqasm} & 25  & 3   & 67.6\% & 482 & 174 \\
        cat\_state\_n22~\cite{cross2021openqasm} & 22  & 2 & 53.2\% & 66 & 21 \\
        rd53\_139~\cite{MDS:2005} & 8 & 5 & 60.9\% & 251 & 140\\
        \hline
        \multicolumn{1}{|l||}{AVG PST(\%)} & \multicolumn{5}{|c|}{51.8\%}  \\
         \hline
         \multicolumn{1}{|l||}{AVG gate count} & \multicolumn{5}{|c|}{214}  \\
         \hline
         \multicolumn{1}{|l||}{AVG CNOT count} & \multicolumn{5}{|c|}{53.1}  \\
         \hline
\end{tabular}
\vspace{-2mm}
\caption{PST results for large benchmark circuits on a 5-qubit ibmq-lima machine (our worst-case scenario).}\label{table:lima}
%\vspace{-6mm}
\end{table*}

%%%%%%%%%%%%%%%%%%%%%%%%%%%%%%%%%%%%%%%%%%%%%%%%%%%%%%%%%%%%%%%%%%%%%%%%%%%%%%%%%%%
\subsection{Iterations vs Shots}\label{sec:shots}. 
While our algorithm seems promising, as demonstrated in Section~\ref{sec:complexity}, there is an underlying issue in current quantum hardware that needs to be carefully addressed. In current systems, the quantum circuit is executed over multiple {\em shots} to attain the probability of success in achieving the correct distribution results~\cite{farhi2014quantum}. However, we observed that when a circuit containing an MM gate is executed using the same strategy, instead of running the circuit until the final operation, the current systems~\cite{ibmq} execute the circuit until the MM gate for multiple shots, attaining the results and continuing with the execution of the remainder of the circuit after that. This will cause the other qubits to be stalled for the number of shots multiplied by the duration of the executed operations on the measured qubit, which can be substantial in practice. Based on the numbers attained from a sample system of the new generation of ibmq, ibmq\_kolkata~\cite{ibmq}, the readout latency is $675\mu S$ and the best available T1/T2 for the qubits is $214\mu S$ (which is the best value  available for this system). If the circuit is stalled for 1000 shots, it means that the other qubits should wait for a minimum of $1000\times 675\mu S= 675 m S$, which is significantly higher than T1/T2 (more than 3000x), causing  the other qubits to become $|0\rangle$ state in the process. None of the current Dynamic  Decoupling~\cite{10.1145/3466752.3480059} techniques can solve an idle period of this magnitude.  

To solve this issue, instead of running a circuit over 1000 shots, we execute the circuit using 1 shot and run this circuit in a for-loop for 1000 iterations. Doing so can solve the issue mentioned above while attaining the final results. It is important to note that, by using this technique, we are not introducing any new significant change in the current systems; rather, we provide a simple method for solving an issue related to current quantum hardware. More specifically, we are encouraging the vendors to add the concept of iterations, which can eliminate the need for unnecessary waiting in the queue. This can be done by tracking the existing job ID and executing the whole circuit instead of a portion of it if the user specifies iteration numbers instead of shots count in the algorithm.

\section{Experimental Evaluation}\label{sec:eval}
%%%%%%%%%%%%
\begin{table*}[ht]
    \centering
    %\vspace{-4mm}
    \begin{tabular}{ |p{3cm}||p{2cm}|p{2cm}|p{2cm}|p{2cm}|p{2cm}|p{2cm}|  }
        \hline
        \multicolumn{7}{|c|}{PST report on ibmq-Kolkata} \\
         \hline
         Circuit Name & Parallel Execution PST & Serial Execution PST & Parallel Execution Gate Count & Serial Execution Gate Count & Parallel Execution CNOT Count & Serial Execution CNOT Count \\
        \hline
        bv\_n14~\cite{cross2021openqasm}   & 26.9\%    & 77.8\% &  285 & 121 & 187 & 13\\
        bv\_n19~\cite{cross2021openqasm} &   21.1\%  & 68.8\%   & 559 & 166 & 426 & 18\\
        wstate\_n27~\cite{cross2021openqasm} & 13.7\% & 56.1\% &  1016 & 593 & 571 & 124\\
        ghz\_state\_n23~\cite{cross2021openqasm}    & 20.5\% & 69.8\% & 307 & 70 & 280 & 22\\
        swap\_test\_n25~\cite{cross2021openqasm} & 43.9\%  & 53.8\%   & 768 & 482 & 484 & 174\\
        cat\_state\_n22~\cite{cross2021openqasm} & 31.3\%  & 73\% & 185 & 66 & 159 & 21 \\
        rd53\_139~\cite{MDS:2005} & 64.5\% & 67.6\% & 245 & 222 & 137 & 111\\
        \hline
        Average & 31.7\% & 66.7\% & 480.7 & 245.7 & 301 & 69\\
        \hline
        \multicolumn{1}{|l||}{AVG PST Gain(\%)} & \multicolumn{6}{|c|}{210.4\% (\~2.1 X)}  \\
        \hline
        \multicolumn{1}{|l||}{AVG gate reduction(\%)} & \multicolumn{6}{|c|}{48.9\% (\~0.5 X)}  \\
         \hline
\end{tabular}
\vspace{-2mm}
\caption{Comparison of our sequential execution and  baseline execution on an ibmq\_kolkata (27-qubit system) simulator.}\label{Table:Kolkata}
\vspace{-6mm}
\end{table*}
%%%%%
In this section, first, we discuss the methodology we adopted in our experiments. We then evaluate our proposal using two different scenarios. Firstly, by using a small quantum hardware, we optimize (serialize) the circuits that {\em cannot}  fit into this hardware using our technique, to show the scalability of our approach. The goal behind experimenting with this scenario is to demonstrate that our approach can be used to execute quantum circuits on quantum hardware with lower qubit capacity. Note that, by default (without our approach) such circuits would not execute on the target (small) quantum hardware. Secondly, using a larger quantum hardware, we evaluate and compare our proposal with original circuits when they can be executed on the quantum hardware. This scenario aims at giving a PST comparison of our proposal against the normal (parallel) and at revealing the improvements we provide due to the factors such as gate reduction.
%%%%%%%%%%%%
\subsection{Methodology}\label{sec:methodology}
%%%%%%%%%%
We evaluate and present our results using two IBMQ systems, namely, ibmq\_lima and ibmq\_kolkata~\cite{ibmq}. ibmq\_lima is a 5-qubit system that has an architecture in a T-like Fig.. For simulating ibmq\_lima, we execute the experiment using FakelimaV2(), the most recent simulator for the latter system supplied by Qiskit~\cite{Qiskit,FakeProvider}. Compared to the earlier version of Fakelima(), FakelimaV2() supports more coherence error-related optimizations like instructions duration, pulse scheduling, etc~\cite{FakeProvider}. For simulated ibmq\_kolkata, we evaluated our results using FakekolkataV2(), which is the fake-backend for the ibmq\_Kolkata hardware. Note that ibmq\_kolkata is a new generation IBMQ system with 27 qubits, each having 1 to 3 links.

While our presented results are based on simulations, we want to emphasize that our comparison is accurate since we  eliminate most of the crosstalk by using serialization. It is because crosstalk occurs when multiple qubits are operated concurrently by different operations (mostly CNOTs) and since most of our quantum circuits can be executed on 2-3 qubits, we are not facing crosstalk in any of the results presented. For our baseline (parallel execution), on the other hand, there may be some crosstalk cases that are ignored; consequently, the baseline results we report can be overestimation in terms of PST. Therefore, our benefits can be expected to be even higher in real quantum hardware. 

Our serialized execution results are reported using 1000 iterations, each containing 1 shot/iteration, as discussed in Section~\ref{sec:shots}. For our baseline results, on the other hand, all the results are reported using 1000 shots per workload. The mapping policy is set to the default mapping that qiskit/qiskit.transpile employs. We report and compare the result by using gate count and PST. Note that gate count is an important metric since it shows the effects of the total gate error. PST is calculated using the following formula (Eval is the correct expected value for the results):

$PST=\sum_{i}^{Eval} \frac{PST[i]}{Count[Eval]}$\newline

%%%%%%%%%%%%%%%%%%%%%%

\vspace{-5mm}
\subsection{Results and Discussion}\label{sec:result}
%%%%%%%%%%%%%%%%%%%%%%
Table~\ref{table:lima} shows our results on a 5-qubit  system for benchmarks as large as 27 qubits. While these (original) circuits clearly cannot run on 5 qubits in a parallel (normal) fashion, we are able to execute them and obtain reliable outputs by using the proposed strategy. Our results indicate that, on average, we are shrinking the size of the circuits tested by 8.87x while achieving an average PST of 51.7\%, which is significant based on the size of the workloads. 

For our experiments, the results are reported on a 5-qubit system, which is the minimum qubit size for the commercially available quantum systems. We use this system to show that i) while prior works cannot execute these algorithms on small hardware, our approach can execute them and achieve results with high reliability, and (ii) there exist some large quantum circuits that benefit from our proposal when targeting even the smallest quantum hardware available. We predict that,  by increasing the qubit size and/or using a newer generation of quantum hardware, our proposal can achieve a better PST for a larger set of workloads.

For Bernstein-Vazirani, we report two results with 14 and 19 qubits. As shown in the Table~\ref{table:lima}, the PST decreases from 51.2\% to 42.2\% for 14 qubits to 19 qubits on ibmq\_lima. While, theoretically, any Bernstein-Vazirani circuit with a given number of qubits can be executed on two qubits~\cite{middleresetIBM} by using the MM and MR gates, fitting larger Bernstein-Vazirani circuits with any number of qubits into 2 qubits does {\em not} always generate a reliable output since it leads to significant coherence error. Therefore, we predict there will be a cap to serialization based on the coherence error of the underlying hardware and the circuit depth. 

\subsection{Sensitivity Analysis on larger systems\protect\footnote{Benchmarks are obtained from QASM~\cite{cross2021openqasm} and Revlib~\cite{MDS:2005}.}}\label{sec:sens}

%%%%%%%%%%%%%%%%%%%%%%%%%%%

In this part, we compare the results of our serialized execution to baseline (parallel) results on a newer generation system. Table~\ref{Table:Kolkata} shows the PST and gate count results with the ibmq\_kolkata system. 
We believe that sequential execution, when applicable, is superior to parallel execution for three major reasons:
\squishlist
    \item Low average number of links on current systems leads to excessive SWAP operations in parallel execution. Our technique can significantly improves this problem.
    \item The new generation of quantum systems is trending toward improving the measurement gate (readout error), which leads to the minimization of our proposal's overhead.
    \item The new generation of the system also has better T1/T2. Therefore, this leads to coherence error decreasing exponentially, which is the main concern for sequential execution.
\squishend  

Therefore, we argue that the current quantum systems are very well suited for sequential execution; in fact, their low average link count is not a good fit for parallel execution, which is the state of the art. Our experimental results indicate that the proposed scheme achieves a 2.1x PST improvement while reducing the number of gates by 48.9\%, compared to running the original circuit on the same system. This is because,  by serializing the execution, we are reducing the number of SWAPs needed to migrate the qubits over the links. Thus, by decreasing the gate error rate, we are able to achieve highly reliable results.
 
Compared to the results reported in Table~\ref{table:lima}, we observe a significant PST improvement with the ibmq\_kolkata system. The reason behind this is that the newer generation IBMQ systems have better system characteristics such as readout latency and T1/T2. Additionally, since the newer generation of quantum hardware leads to lower readout latency, we expect the effectiveness of our approach to be even higher in future systems. 

%%%%%

%%%%%
\vspace{-3mm}
\section{Conclusion} \label{sec:conclusion}
%%%%%%%%%%%%%%%%%%
In this paper, we presented an approach for sequentially executing quantum circuits and resizing them into the smallest qubit count necessary to fit them into a small-sized system using the MM/MR gates.
We demonstrated the correctness of our proposed method and provided a complexity analysis demonstrating that it operates in $\mathcal{O}(mn^2\log{}n)$ time. We also showed its scalability by choosing the smallest system with the largest impact on coherence error, which is the primary concern in sequential execution, and reported the appropriate level of reliability for a large fraction of the (large) benchmark circuits offered by QASM~\cite{li2020qasmbench}.

We suggested the notion of iterations number over the prior concept of shots number in order to reduce coherence error and deliver reliable results on a small worst-case system for large benchmark circuits, as opposed to no results at all (compilation size error).
Sequential execution can possibly compensate for quantum devices' lack of memory to a certain degree by resizing the circuits and allowing small chunks of circuit to execute on processing qubits. 
We also showed, via simulations, that, on a modern NISQ system with 27 qubits, owing to the reduction in the number of links in the current NISQ~\cite{ibmq} designs, our proposed sequential execution can boost reliability by a factor of two (avg. 2.1x PST improvement) and reduce gate counts by almost half by minimizing the SWAP operations. 

%%%%%
\section{Future Work}
\label{sec:future}
%%%%%%
Our approach acts on the basis of dependencies extracted from a DAG to decrease the circuit size via serial/sequential execution while preserving the circuit's structure.
As a future direction, one can explore an approach to obtain true dependencies in polynomial time and feed them as input to our algorithm for additional size reductions. However, we plan to explore a potentially better strategy, which is to construct circuits in early stages (e.g., Pualihedral IR~\cite{li2022paulihedral}) such that the DAG dependencies do not impact the size reduction, as shown in Fig.~\ref{fig:BVN} (BV-n circuit). In this way, the circuit is constructed in an ideal fashion from the viewpoint of serial execution, and we can employ our algorithm while preserving the circuit's structure and changing just its execution method. 

Future quantum hardware can enhance qubits connectivity (links), owing to anticipated improvements in fidelity and/or reliability. We suggest developing techniques for combination of sequential execution and current parallel execution based on the architecture of the systems in order to increase the circuit's reliability and reduce its gate count. We also recommend mapping policies improvements and the development of the dynamic decoupling (DD) strategies for sequential execution of circuits to further boost reliability, similar to the techniques that have been established for the existing method of execution~\cite{liu2021qucloud,ohkuracrosstalk,tannu2019not}. Additionally, based on the observations in Section~\ref{sec:shots}, we also encourage hardware designers to incorporate support for the concept of iterations in addition to the shots. 

\section{Disclaimer }\label{disclaimer}

We are aware of two works concurrent with our paper. These two works have recently appeared in arXiv, and aim to reduce qubit requirements via qubit reuse. One of these works~\cite {mmmrOctober} (dated October 14th, 2022) was applied on top of ion trap machines with all-to-all connections in an effort to reduce resource usage. For smaller circuits, a SAT-based technique is used to determine optimal utilization, whereas a heuristic method is employed for larger circuits. However, this study did not consider one of the main advantages of our strategy, which is to minimize the number of SWAP operations. In addition, our compiler-based approach  can outperform the dynamic algorithm in \cite{mmmrOctober} through our greedy strategy, which has been proven to be optimal in this paper. The other work~\cite{hua2022exploiting} (dated November 3rd, 2022) explores  a similar approach to fit a program in the selected hardware. Our paper differs from that work in that, we prove that our algorithm really minimizes the circuit in a polynomial amount of time, whereas the mentioned work does not, and We want to emphasize these two arXiv papers are concurrent works to ours, and  our proposal had been submitted to (and rejected in) an earlier conference deadline (August 1st, 2022), predating both of these arXiv submissions. With the exception of this subsection~\ref{disclaimer} we submitted the paper as it looked by August 1st, 2022 ignoring all the later updated versions.

%%%%%%

%%%%%%% -- PAPER CONTENT ENDS -- %%%%%%%%

\pagebreak
%%%%%%%%% -- BIB STYLE AND FILE -- %%%%%%%%
\bibliographystyle{IEEEtranS}
\bibliography{refs}
%%%%%%%%%%%%%%%%%%%%%%%%%%%%%%%%%%%%

\end{document}